\documentclass[%
 reprint,
 amsmath,amssymb,
 aps,
]{revtex4-2}

\usepackage{graphicx}
\usepackage{dcolumn}
\usepackage{bm}
\usepackage{color}
\usepackage{ulem}

\begin{document}

\preprint{APS/123-QED}







\title{High temperature superradiant phase transition in novel quantum  structures   with complex network interface}

\author{A.Yu. Bazhenov, M. Nikitina, and Alexander Alodjants}
\email{alexander\_ap@list.ru}
\affiliation{ITMO University, St. Petersburg, 197101, Russia}

\begin{abstract}
In the present work we propose a novel quantum material  concept, which enables super- and/or ultrastrong interaction of two-level systems with the photonic field in  a  complex network. Within the mean field approximation we examine phase transition to superradiance that results in two excitation (polariton) branches and is accompanied by the appearance of non-zero macroscopic polarization of two-level systems. We characterize the statistical properties of networks by the first, ${\langle}k{\rangle}$, and second normalized, $\zeta\equiv{\langle}k^2{\rangle}/{\langle}k{\rangle}$,  moments for node  degree distribution. We have shown that the Rabi frequency is  essentially enhanced due to the topology of the network within the anomalous domain where ${\langle}k{\rangle}$ and $\zeta$ sufficiently grow. The multichannel (multimode) structure of matter-field interaction leads superstrong coupling that provides primary behavior of the high temperature phase transition.  The results obtained pave the way to design new  photonic and polaritonic circuits, quantum networks  for efficient processing quantum information at high (room) temperatures.     
\end{abstract}


\maketitle

The enhancement of matter-field interaction represents a keystone problem for current quantum technologies,  which use photonic field as a carrier of quantum information, see e.g.  \cite{Alodjants}. Currently available materials possess  strong and super(ultra)-strong coupling of matter and field \cite{Kockum,Meiser}. The strong coupling presumes  observation of periodic energy  exchange between two-level systems (TLS), for example, atoms, excitons,  Cooper-pair boxes,   etc. and quantized field \cite{Miller, Skolnick,Blais}. 
The frequency of oscillations is the Rabi-splitting  frequency, which depends on matter-field interaction strength and is characterized by the Rabi-splitting effect exhibiting so-called $\sqrt{N}$ scaling law for collective matter-field coupling parameter $g=\sqrt{N}g_0$;  $2g_0$  is  the single-photon Rabi frequency, and $N$ is a number of TLSs interacting with a single mode cavity field \cite{Miller}.  At the  core of such a behaviour there is cooperative (coherent) interaction of TLSs  with some privileged high Q-cavity irradiation mode. In this case,  the Dicke spin behaves as a giant quantum oscillator effectively coupled  with a singe quantized mode with coupling strength $g$.
 
In theory, strong coupling regime appears if  $g_0$ prevails over spontaneous decay, dephasing rates  and cavity losses \cite{Miller};  current quantum technologies allow for the observation of the vacuum Rabi-splitting achieved with a single TLS \cite{Miller, Skolnick}. However, in practice the improvement of collective matter-field coupling  parameter  represents an imperative task for quantum information,  which explores various "hardware" platforms for quantum information processing and transmission, cf. \cite{Schneeweiss}. The efforts are directed to obtain (collective) superstrong  coupling regime when coupling between TLSs and a quantized  mode behaves  strongly enough during one round trip in the cavity, cf. \cite{Meiser, Johnson,Kuzmin}.  A possible solution to this problem may be the realization of collective  interaction between TLSs and the multimode quantized field \cite{Neereja, Schneeweiss}. Notably, a large cavity length (optical fiber)  contributes to the achievement of superstrong coupling for cold ensemble of two-level atoms with the optical field.

In this work, we suggest a new  quantum  artificial  material  concept, which mimics a complex networks architecture  and enables to create superstrong coupling between a collective spin and cavity modes established by the  network topology, cf. \cite{Bazhenov, Samura, Lepri}.  It is known that hubs occurring in complex networks provide much faster and robust information spread \cite{Barabasi1}. We demonstrate how the network architecture influences the Rabi-splitting effect and occurrence of high temperature phase transition to superradiant (SR) state. 
Since early seminal works \cite{Dicke, Hepp, Hioe}  superradiance and SR phase transitions have been intensively studied for various physical systems of radiation and matter interaction, see e.g. \cite{Vyas} and references therein. In this sense the results obtained serve a useful prerequisite for new  experimental activity in this field.

\begin{figure*}[ht]
\includegraphics[width=\linewidth]{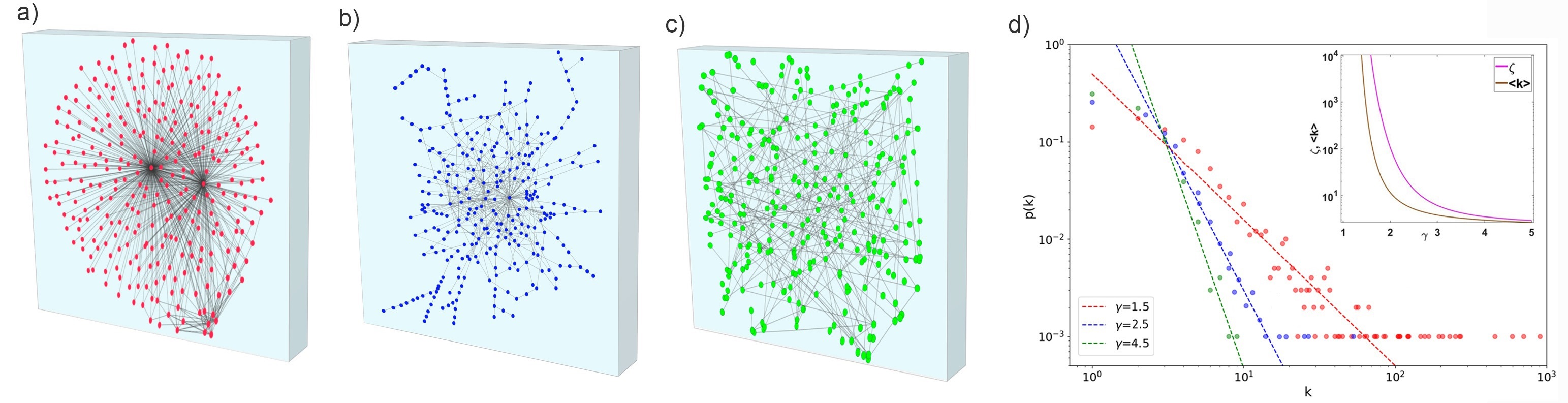}
\caption{Samples of quantum artificial microstructures with interfaces possessing power law $p(k)\propto k^{-\gamma}$ node degree   distribution for photonic channels  with  a) $\gamma = 1.5$,  b) $\gamma = 2.5$  and  c) $\gamma =4.5$, respectively.  d)  dependence of $p(k)$ on $k$ given in log plots.  The inset  in d) shows dependence ${\langle}k{\rangle}$ and $\zeta$ on $\gamma$ in logarithmic scale  for the network with number of nodes $N=300$ and $k_{min}=2$.}
\label{FIG:Networks}
\end{figure*}

The model of quantum structure  we establish here consists of $N$ localised TLSs, which occupy graph nodes at temperature $T$, as shown in Fig. \ref{FIG:Networks}. Any $j$-th node possesses some number of edges,  which  link the node  with  other ones. The number of edges, which are incident to any $j$-th node, may be characterized by the node degree $k_j$.  In this work we consider the case when graph architecture in Fig. \ref{FIG:Networks} may be designed by two-level  quantum dots  (QD) located on the 2D photonic crystal microstructure, which provides cavities and waveguides \cite{Skolnick, Notomi}. In particular, links between the nodes represent photonic crystal waveguides.

The Hamiltonian $H$  that describes the interaction of  a photonic field with TLSs for the graph structure in Fig.~\ref{FIG:Networks} we  establish  as
\begin{widetext}
\begin{equation}\label{Hamiltonian}
\begin{gathered}
H = \hbar\sum\limits_i^N{\frac{{{\omega _{0,i}}\sigma _i^z}}{2}} + \hbar\omega _{ph}\sum\limits_v^{k_i}{a_v^\dag }a_v + \frac{  \hbar g }{\sqrt N }\sum\limits_i^N\sum\limits_v^{k_i} {\left( {a_v\sigma _i^ + +a_v^\dag\sigma _i^ -} \right)},
\end{gathered}
\end{equation}
\end{widetext}
where $a_v (a_v^\dag)$ is annihilation (creation) operator for the $v$-th waveguide field, 
$\sigma_i^z$ 
is an operator of population inversion for $i$-th TLS, \({\omega _{0,i}}\) is the resonant frequency of the transition TLSs from the ground to the excited state and $g$ 
is a strength of the interaction of TLSs with photonic field possessing frequency $\omega_{ph}$. Thereafter we use units when $\hbar=k_B=1$ is taken for Planck and Boltzmann constants, respectively.
The Hamiltonian  ~\eqref{Hamiltonian} commutes with operator of excitation $N_{ex,i} =\sum\limits_i^N {\frac{{\sigma _i^z}}{2}} + \sum\limits_v^{k_i}{a_v^\dag }a_v$ of the $i$-th node.  
Then, we suppose  the photonic modes  to be in coherent states ${\left| \alpha \right\rangle}$, defined by $a_v{\left| \alpha_v \right\rangle}=\alpha_v{\left| \alpha_v \right\rangle}$, where  $\alpha_v$ is taken real and equal to each other for all nodes, i.e. we assume  $\Lambda\equiv\alpha_v$ in ~\eqref{Hamiltonian}. In particular, $\alpha$ represents a variational (order) parameter. 
Since the coupled matter-field system is open, it is more suitable to use the grand canonical ensemble approach that implies non-zero chemical potential $\mu$. In particular,  we can use in ~\eqref{Hamiltonian}  the following substitutions  $\Omega_{0,i}=\omega_{0,i}-\mu$ and $\Omega_{ph}=\omega_{ph}-\mu$, which account the grand canonical approach to the problem, cf. \cite{Baznehov2}. 
In this limit we can represent the Hamiltonian ~\eqref{Hamiltonian} as:
\begin{eqnarray}\label{Hamiltonian2}
H = \sum\limits_i^N {\frac{{{\Omega _{0,i}}\sigma _i^z}}{2}} + {\Omega _{ph}}{\langle}k{\rangle}\Lambda^2 + \frac{{g \Lambda }}{{\sqrt N }}\sum\limits_i^N {\left( {\sigma _i^ + +\sigma _i^ -} \right)}. \, \end{eqnarray}

In this work, we are interested in excitation density  $\rho\equiv\frac{1}{N}{\langle}\sum\limits_i^N N_{ex,i}{\rangle}$,  which represents a normalized average total number of excitations and  may be obtained  in the form
\begin{equation}\label{densities}
\rho=\frac{{\langle}k{\rangle}\Lambda^2}{N} + \frac{1}{{2}} {S_z}.\
\end{equation}
In \eqref{densities} $\Lambda^2 \equiv N_{ph}$ is an average photon number in the network structure, and ${S_z} = \frac{1}{N}\sum\limits_{i = 1}^N {\langle\sigma _i^z\rangle} $ is an average collective population imbalance. 
In the framework of mean-field theory for  partition function $Z(N,T)=Tr(e^{-\beta H})$, which explores  ~\eqref{Hamiltonian2},  we obtain  
\begin{widetext}
\begin{equation}\label{partition3}
Z(N,T) = e^{-\beta \Omega_{ph}{\langle}k{\rangle}\Lambda^2}\prod\limits_i^N{2cosh\left(\frac{\beta}{2}\sqrt{\Omega_{0,i}^2+4\frac{g^2k_i^2}{N}\Lambda^2}\right)}, 
\end{equation}
\end{widetext}
where $\beta\equiv1/T$ is reciprocal temperature. We assume that the number of nodes  is large enough, $N\gg1$, and the  network structure admits the transition to  continuous degree distribution $p(k)$. This  transition  requires  replacement $\frac{1}{N}\sum\limits_{i}{...} \rightarrow\int\limits^{k_{max}}_{k_{min}}{...p(k)dk}$, where $k_{min}$ and $k_{max}$ are the minimal and maximal values of node degree $k$,  cf. \cite{Bazhenov}. In this case with the help of  \eqref{partition3} we obtain  
\begin{subequations}\label{basic}
\begin{align}
\begin{split}
{ \Omega_{ph}}= \frac{g^2}{{\langle}k{\rangle}}\int\limits_{  k_{min} }^{k_{max}} {\frac{k^2}{\Gamma}\tanh \left[ {\frac{\beta \Gamma}{2}} \right]} p(k)dk; \end{split}\label{system1}\\
\begin{split}
\rho = {\langle}k{\rangle}\frac{\Lambda^2}{N} - \frac{1}{2}\int\limits_{ k_{min} }^{k_{max}} {\frac{{ {{\Omega }_0}}}{\Gamma}\tanh \left[{\frac{\beta \Gamma}{2} } \right]} p(k) dk,\end{split} \label{system23}
\end{align}
\label{system2}
\end{subequations}
where we define $\Gamma\equiv\sqrt {{{\Omega }_0^2} + 4\frac{k^2}{N}{g^2\Lambda ^2}}=\sqrt {{{\Omega }_0^2} + 4k^2g_0^2\Lambda ^2}$ and propose that all  TLSs are identical to each other setting in  \eqref{basic} $\Omega_{0,i}=\Omega_0$. 

Set of   \eqref{basic} characterizes basic features of the order parameter $\Lambda$  (\eqref{system1}) and excitation density  $\rho$  (\eqref{system23}) at thermal equilibrium. We solve them self-consistently searching the chemical potential $\mu$  for various network topology.   
In particular, we examine networks topology, which is determined by power-law distribution function  $p(k)$ defined as (cf. \cite{Barabasi1}) 
\begin{eqnarray}\label{powere}
p(k)=\frac{(\gamma-1)k_{min}^{\gamma-1}}{k^{\gamma}},
\end{eqnarray}
where $\gamma$ is a degree exponent, $k_{min}$ is the smallest degree for which \eqref{powere} holds. 
The network with distribution  ~\eqref{powere} obeys the normalization condition 
$\int\limits^{+\infty}_{k_{max}}{p(k)dk}=\frac{1}{N}$, which implies that the network with $N$ nodes possesses more than one node with $k>k_{max}$. From ~\eqref{powere} we obtain $k_{max}=k_{min}N^{\frac{1}{\gamma-1}}$. 

We characterize the statistical properties of the networks by the first (${{\langle}k{\rangle}}$) and second (${{\langle}k^2{\rangle}}$) moments for the degree distribution, which are defined as:
\begin{eqnarray}\label{moments}
{{\langle}k^m{\rangle}}=\int\limits^{k_{max}}_{k_{min}}{{k^m}p(k)dk}, \; \; m=1,2.
\end{eqnarray}
Below we exploit the parameter $\zeta=\frac{{\langle}k^2{\rangle}}{{\langle}k\rangle}$, that determines the basic statistical properties of the chosen network. Degree exponent  $\gamma$ defines three different domains, which determine seminal   features for  ${{\langle}k{\rangle}}$ and $\zeta$ in  anomalous ($1<\gamma < 2$), scale-free ($2<\gamma < 3$), and random ($\gamma>3$) regimes, respectively.  The properties of  networks possessing distribution~\eqref{powere} for $\gamma=2$ and $\gamma=3$ should be calculated separately.

In Fig.~\ref{FIG:Networks}(a)-(c) we represent samples with the network topology, which   corresponds to   anomalous (Fig.~\ref{FIG:Networks}(a)), scale-free (Fig.~\ref{FIG:Networks}(b)), and random (Fig.~\ref{FIG:Networks}(c)) regimes, respectively.  Key-stone behaviour of such a networks occurs due to the presence  of hubs, which are clearly seen as several  points located in the right corner of  Fig.~\ref{FIG:Networks}(d).  The largest hub is described by degree $k_{max}$ and called a natural cutoff. The  network in anomalous regime exhibits maximal number of hubs, see Fig.~\ref{FIG:Networks}(a). Contrary, in the random regime the number of hubs sufficiently vanishes, see the green curve in  Fig.~\ref{FIG:Networks}(d).

\begin{figure}[ht]
\begin{minipage}[h]{0.8\linewidth}
\center{\includegraphics[width=1\linewidth]{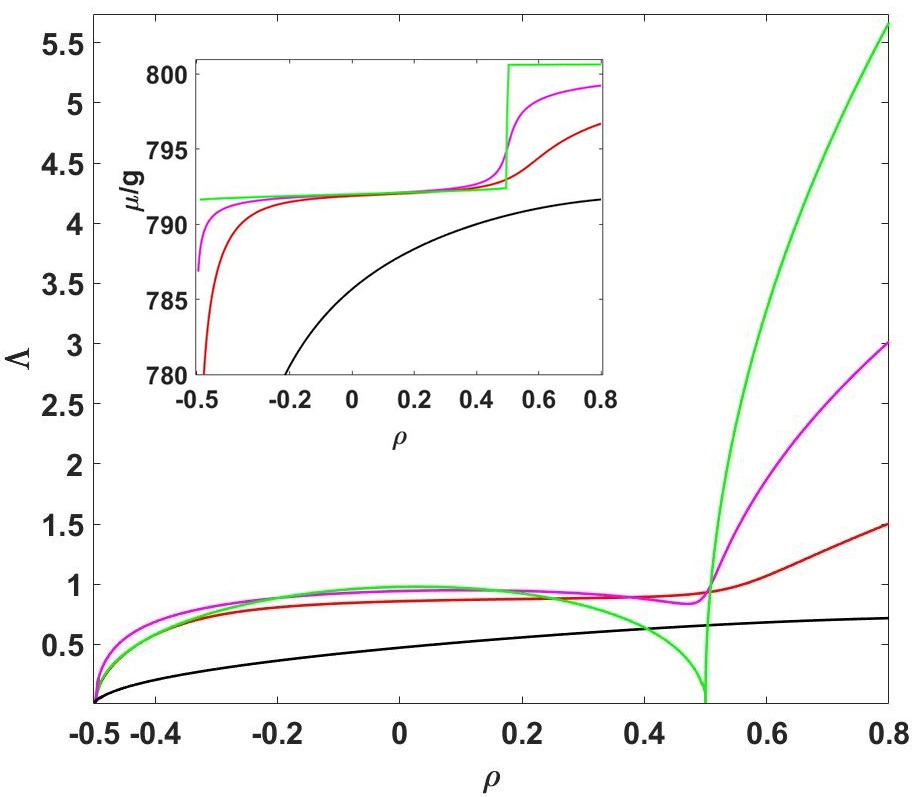} \\ a)}
\end{minipage}
\hfill
\begin{minipage}[h]{0.8\linewidth}
\center{\includegraphics[width=1\linewidth]{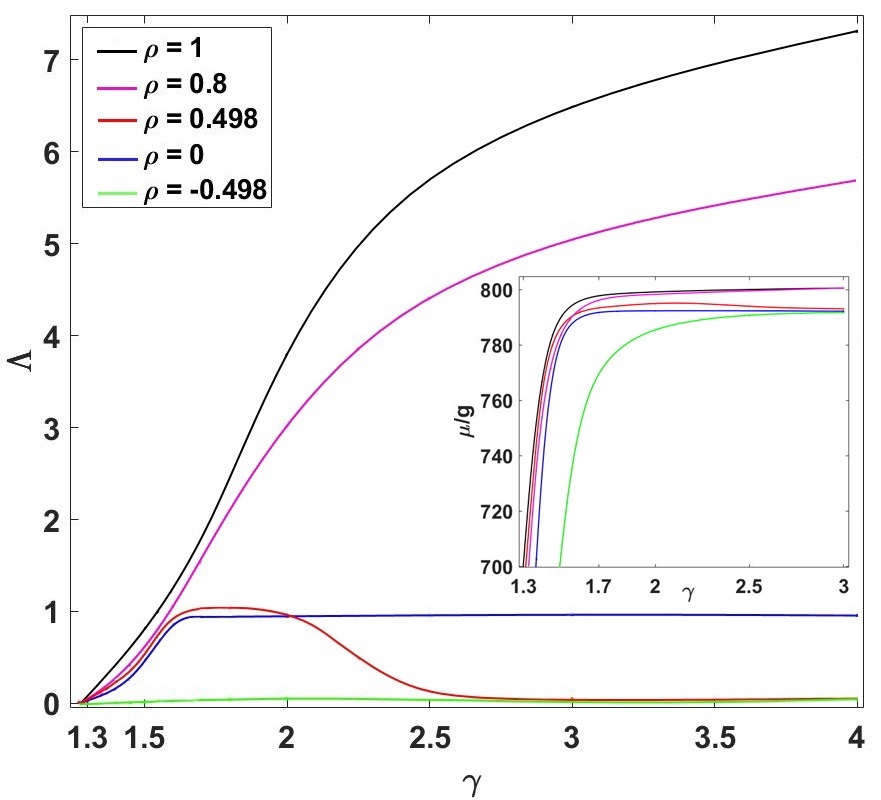} \\ b)}
\end{minipage}

\caption{Zero temperature dependence of the order parameter $\Lambda$ and normalized chemical potential $\mu/g$ (inset) as the function of (a) the excitation density $\rho$ for $\gamma = 1.5$ (black), $\gamma = 1.7$ (red), $\gamma = 2$ (purple), $\gamma = 4$ (green) curves, respectively, and (b) degree exponent $\gamma$ for the network of $N=300$ QD's.
The other parameters  are: $\Delta/g = 9$, $\omega_0/g = 792$,  $k_{min}=2$, $g\equiv g_0\sqrt{N}\simeq1836$ $\mu eV$.} 
\label{FIG:2}
\end{figure}

Self-consistent numerical solutions of \eqref{basic}  demonstrate that order parameter $\Lambda$ monotonously increases with $\rho$ in the presence of resonance, i.e. at  $\Delta=0$. 
In Fig.~\ref{FIG:2}(a) we establish these solutions   for $\Lambda$ and the chemical potential, $\mu$,
 as the function of excitation density $\rho$ for  positive   detuning  $\Delta\equiv \omega_{ph}-\omega_0$  in  limit $T \rightarrow 0$.  
 To be more specific,  we examine InGaAs/GaAs QD's incorporated into photonic crystal network environment for numerical estimations in Fig.~\ref{FIG:2}. In particular, achieved in \cite{Skolnick} a strong coupling regime implies $g_0=106$ $\mu eV$ for $\omega_{ph}/2\pi\simeq353$ $THz$ ($\lambda=850$ $nm$).
  In low excitation density from Fig.~\ref{FIG:2}(a) we can see that $\Lambda \rightarrow 0$,  $S_z\simeq-1$,  which  implies   $\rho\simeq -0.5$; the ensemble of TLSs is inversion-less. The behaviour of chemical potential is easy to obtain analytically in the limit of  $\gamma>3$ when the network  (see Fig.~\ref{FIG:2}(c)) resembles regular network features with ${\langle}k{\rangle}\to k_{min}$.   
In this case from \eqref{basic} we obtain 
\begin{eqnarray}\label{muSR0}
\mu_{1,2}\simeq\mu_0 \pm \frac{1}{2}\sqrt{\Delta^2  - 8g^2{\langle}k{\rangle} \left(\rho-\frac{{\langle}k{\rangle}}{N}\Lambda^2\right)},
\end{eqnarray}
where $\mu_0\equiv(\omega_{ph}+\omega_{0})/2$. 
\eqref{muSR0}   establishes upper ($\mu_1$) and lower ($\mu_2$)  excitation branches for coupled matter-field states, cf. \cite{Hui}. At $\rho=0$ TLS  undergoes saturation when the number of particles at the ground and upper levels are equal to each other, $S_z\simeq0$. 
The limit $\rho>0$  establishes inversion occurring in  TLS, which is maximal at  $\rho= 0.5$ ( $S_z=1$).   Notably, in the presence of large detuning $\Delta$ TLS undergoes structural transition to another (parametrical) type of excitations, which are inherent to  strong amplification of irradiation occurring at  $\rho>0.5$,  cf. \cite{Baznehov2}. 
The condition to achieve such a transition may be found from \eqref{muSR0} and represented as $|\Delta| > 2g \sqrt{{\langle}k{\rangle}}$. In Fig.~\ref{FIG:2}(a) this  inequality  fully violets for the black curve, which corresponds to an increasing value of  ${\langle}k{\rangle}$ with $\gamma=1.5$. 
In particular, for large enough detuning $\Delta$ from \eqref{muSR0} we obtain $\mu_1\approx \omega_0+\Delta=\omega_{ph}$,  $\mu_2\approx \omega_0$,  respectively. At  $\rho=0.5$ the green curve in the inset to Fig.~\ref{FIG:2}(a) establishes abrupt transition for the chemical potential from the lower polariton branch to the upper one due to occurring full TLS inversion ($\Lambda\simeq 0$); $\mu_1-\mu_2\simeq \Delta$ is  the gap between two branches in Fig.~\ref{FIG:2}(a). The dependence of  $\Lambda$ on degree exponent $\gamma$ is shown in Fig.~\ref{FIG:2}(b) and completely agrees with the order parameter features given in  Fig.~\ref{FIG:2}(a). At  large $\gamma$ the chemical potential  (see the inset to Fig.~\ref{FIG:2}(b)) clearly demonstrates  asymptotic behaviour that is relevant to the gap  between the polariton branches. For the upper (the black and magenda curves in Fig.~\ref{FIG:2}(b)) the excitation density is positive and the order parameter grows as $\Lambda\simeq\sqrt{N\rho/{\langle}k{\rangle}}$. For inversion-less TLS $\rho$ is negative and $\Lambda$ vanishes, see the red and green curves in Fig.~\ref{FIG:2}(b).  

Let us focus on non-zero temperatures for the system in Fig.~\ref{FIG:Networks}. The behaviour of chemical potential is easy to obtain analytically in high temperature limit $\beta \rightarrow 0$.  In this case from \eqref{basic} we obtain, cf. \eqref{muSR0}.
\begin{eqnarray}\label{muSR00}
\mu_{1,2}=\mu_0 \pm \frac{1}{2}\sqrt{\Delta^2  - 8g^2\zeta \left(\rho-\frac{{\langle}k{\rangle}}{N}\Lambda^2\right)}.
\end{eqnarray}
\eqref{muSR00} is valid for any $\gamma>1$ and constitutes one of the main results of this work. From  \eqref{muSR00}, due to dependence on $\zeta$,  it is evident that Rabi-splitting frequency (last term in \eqref{muSR00}) grows for the anomalous regime of a degree exponent, cf. inset in Fig.~\ref{FIG:Networks}(d). Thus, we are able to improve cooperative interaction of TLS with the quantized field by choosing  appropriate $\gamma$,  which dictates the circuit topology shown in Fig.\ref{FIG:Networks}.  Notice that at  large $\gamma$ (see the inset in Fig.~\ref{FIG:Networks}(d)) $\zeta\approx {\langle}k{\rangle}$ and \eqref{muSR0} practically coincides with  \eqref{muSR00}.  

The phase transition to SR state may be recognized from \eqref{basic} setting $\Lambda=0$. An explicit result may be obtained for low branch polaritons  at $\Delta=0$. For  critical  temperature of phase transition  $T_c$ we obtain   
\begin{eqnarray}\label{TcSR}
T_c=\frac{\sqrt{-8g^2\zeta\rho}}{4 \tanh^{-1}(2\rho)}.
\end{eqnarray}
\eqref{TcSR} represents another important result of this work.
As can be seen from \eqref{TcSR}, for $\rho\simeq 0$, the critical temperature $T_c \rightarrow \infty$.  This limit corresponds to the saturation of TLSs.
From  \eqref{TcSR} it is evident that SR phase transition temperature $T_c$  may be very  high  even in a low excitation density limit due to network statistical properties, which are accounted in \eqref{TcSR}  by $\zeta$-parameter, see the inset in Fig.\ref{FIG:Networks}(d).  
 Noteworthy, in annealed networks with Ising-type spin-spin interactions the temperature  is proportional to  $\zeta$-parameter in the critical point of phase transition, cf.\cite{Bianconi0, Bazhenov}.   

For a given temperature $T$, and $-0.5<\rho <0$ \eqref{TcSR} presumes critical value $\zeta_c=-\frac{2 T^2 [\tanh^{-1}(2\rho)]^2}{\rho g^2}$, which  establishes a critical features of the network when SR phase transition occurs. In particular, SR state exists for $\zeta\geq\zeta_c$.
The behaviour of the order parameter within the domain of  large enough  $\zeta$ and  high temperatures  may be obtained from \eqref{system23}   in the form 
\begin{eqnarray}\label{Lambda}
\Lambda\simeq\sqrt{\frac{N_c x_c}{{\langle}k_c{\rangle}}\left(\frac{x}{x_c}-1\right)},
\end{eqnarray}
where $x\equiv\frac{\beta g}{4}\sqrt{-2\zeta\rho}$ is the combination of key-parameters of the coupled matter-field system, and $x_c$ assigns this combination in the phase transition point for  $\zeta_c$.
\eqref{Lambda} establishes a vanishing  SR photon number in the vicinity of the phase transition point.  

The analysis of the system in Fig.\ref{FIG:Networks} in the presence of losses and decoherence  should be performed separately. In the framework of Heisenberg-Langevin formalism self-consistent equations can be obtained for average (coherent) field  amplitude $\Lambda$ and collective  polarization  ${S_-} = \frac{1}{\sqrt N}\sum\limits_{i = 1}^N {\langle\sigma _i^{-} \rangle}$ in the low excitation density limit, which  implies that TLSs are inversion-less $S_z\simeq -1$ ($\rho\simeq-0.5$). By  \eqref{Hamiltonian} we  obtain
\begin{subequations}\label{Set}
\begin{align}
\begin{split}
\frac{d\Lambda}{dt}=-(i\omega_{ph}+\kappa){\langle}k{\rangle}\Lambda-ig S_{-}; \end{split} \label{systemf1}\\
\begin{split} 
\frac{dS_{-}}{dt}=-(i\omega_{0}+\Gamma)S_{-} -ig \Lambda {\langle}k{\rangle},\end{split}\label{systemf2} 
\end{align}
\end{subequations}
where $\kappa$ and $\Gamma$ characterize photon losses and depolarization rates, respectively, cf. \cite{Hui}. The stationary solutions of \eqref{Set} lead to eigen-energies $\mu_{1,2}$ in the form
\begin{widetext}
\begin{eqnarray}\label{muDis}
\mu_{1,2}=\mu_0 -i\frac{1}{2}(\kappa+\Gamma) \pm\frac{1}{2}\sqrt{(\Delta-i(\kappa-\Gamma))^2+4g^2{\langle}k{\rangle}}.
\end{eqnarray}
\end{widetext}
\eqref{muDis} is nothing else as energies of the upper and low polariton branches obtained  within mean-field theory at low excitation density limit, possessing $\zeta\approx {\langle}k{\rangle}$, cf.  \eqref{muSR0}. Notably, average degree ${\langle}k{\rangle}$ increases within the anomalous regime where degree exponent $\gamma$ diminishes, see the inset the inset in Fig.\ref{FIG:Networks}. Noteworthy, large values of ${\langle}k{\rangle}$ enable to represent a collective superstrong coupling condition for interaction of TLSs with the quantized field in the structure (Fig.~\ref{FIG:Networks}) in the form 
\begin{eqnarray}\label{strong}
g{\langle}k{\rangle}^{1/2}>\Gamma, \kappa, \omega_{FSR},  
\end{eqnarray}
where $\omega_{FSR}=c/2L$ is  a free spectral range  of a cavity, $L$ is cavity length, cf. \cite{Meiser}. 
Physically, \eqref{strong} possesses sufficient  enhancement of cooperative matter-field interaction due to the dependence on average degree ${\langle}k{\rangle}$ and existence of different interaction channels and hubs as a result. For large enough ${\langle}k{\rangle}$ condition \eqref{strong}  may be fulfilled during direct spreading of irradiation in the microstcurture and without its circulation, i.e. for one round trip and relatively small $L$, cf. \cite{Meiser, Johnson}. Moreover, one can expect that the microstuctures with the anomalous network domain may allow to obtain the ultrastrong coupling regime when $g{\langle}k{\rangle}^{1/2}$ approaches bare state frequency $\omega_0$, or $\omega_{ph}$. In the optical frequency domain the ultrastrong coupling regime currently represents a challenging problem, cf. \cite{Kockum}.

In Conclusion, we propose a new quantum  material concept, which considers the complex network structure possessing super(ultra)-strong cooperative coherent interaction of  two-level systems occupying network nodes with the quantized optical field. The superstrong regime occurs due to the special network architecture that provides simultaneous occupation of interacting  quantized field many channels (network edges) of the structure. We predict $\sqrt{{\langle}k{\rangle}}$ law for collective matter-field strength behaviour which supports giant  enhancement of matter-field interaction especially  within anomalous domain of the  network structure. We examine the problem of superradiant phase transition for the quantized field amplitude, which  occurs under the interaction of TLSs with irradiation in the quantum material structure. 
It is important that the temperature of phase transition may be very high in a low excitation density limit within the same  anomalous domain of the network structure when statistical properties of the network embodied in the first and second moments of degree  significantly  increase. The  results obtained open new opportunities in network  processing of quantum information as well as observing and exploring Bose-Einstein condensation of  polaritons occurring in network-like microstructures  at high enough temperatures.

\textbf{Funding} We acknowledge  financial support from Russian Priority 2030 program.

\textbf{Disclosures} The authors declare no conflicts of interest.

\bibliography{biblio}


\end{document}